\documentclass[11pt,letterpaper]{article}
\usepackage{graphicx}
\usepackage{amsmath,amsfonts,amssymb}
\usepackage[bf]{caption2}
\begin{document}

\begin{center}
\large{ \textbf{Monitoring of Orientation in Molecular Ensembles
by Polarization Sensitive Nonlinear Microscopy}} \\
\small{V\'eronique Le Floc'h, Sophie Brasselet\footnote{Author to
whom correspondence should be addressed. Email:
sophie.brasselet@lpqm.ens-cachan.fr}, Jean-Fran\c{c}ois Roch, Joseph Zyss\\
Laboratoire de Photonique Quantique et Mol\'eculaire (UMR 8537),\\
Institut d'Alembert (IFR 121), \'Ecole
Normale Sup\'erieure de Cachan\\
61, avenue du Pr\'esident Wilson, 94235 Cachan Cedex, France}
\end{center}

We present high resolution two-photon excitation microscopy
studies combining two-photon fluorescence (TPF) and second
harmonic generation (SHG) in order to probe orientational
distributions of molecular ensembles at room temperature. A
detailed polarization analysis of TPF and SHG signals is used in
order to unravel the parameters of the molecular orientational
statistical distribution, using a technique which can be extended
and generalized to a broad variety of molecular arrangements. A
polymer film containing molecules active for TPF and/or SHG
emission is studied as a model system. Polarized TPF is shown to
provide information on specific properties pertaining to
incoherent emission in molecular media, such as excitation
transfer. SHG, being highly sensitive to a slight departure from
centrosymmetry such as induced by an external electric field in
the medium, complements TPF. The response of each signal to a
variable excitation polarization allows investigation of molecular
behavior in complex environments which affect their orientations
and interactions.

\section*{I. Introduction}
Multi-photon microscopy has amply demonstrated its assets in the
study of broad variety of physical and biological phenomena. Among
the effects arising from two-photon excitation, two-photon
fluorescence (TPF) has been at the focus of much attention for
many reasons.\cite{generalite} Contrary to linear fluorescence,
for which background noise rejection requires the implementation
of confocal detection, TPF exhibits an intrinsic high spatial
resolution due to the built-in quadratic power dependence on the
excitation intensity. In addition, the use of infrared excitation
minimizes both optical damage and background scattering in complex
samples such as living cells.\cite{Webb} This method has shown
significant advantages in imaging of complex biological
environments down to the single molecule level.\cite{MertzXu} In
parallel, developments in molecular engineering have lead to the
design of chemical probes exhibiting very high two-photon
absorption cross sections.\cite{2PhotonDyes,Blanchard} Such
results have triggered important advances in the field of
two-photon imaging and other processes such as optical limiting.
Finally, multi-photon microscopy techniques have been extended to
a larger variety of nonlinear optical processes such as
three-photon induced fluorescence, as well as second- and
third-harmonic frequency generation both in far-field and
near-field microscopies.\cite{Silberberg,Schins,Jakubczyk,Novotny}

Second harmonic generation (SHG) is the result of coherent
emission from anharmonic oscillators upon two-photon excitation.
At macroscopic scale, SHG emission va\-nishes in centrosymmetric
media, since the $\chi^{(2)}$ related nonlinear susceptibility is
an odd-rank tensor. In molecular media, non-centrosymmetric
ordering is traditionally obtained by electric field poling of
intrinsically non-centrosymmetric molecules, such as dipolar
$\pi$-con\-ju\-ga\-ted systems functionnalized with adequate
acceptor and donor groups.~\cite{Zyss} As SHG is highly sensitive
to even a slight departure from centrosymmetry in molecular media,
this effect has been widely used to study surface and interface
properties in physics and chemistry, with monolayer
sensitivity.\cite{Shen,Eisenthal} SHG signals have also been
measured in artificial vesicles,\cite{Mertzvesicles} live
cells,\cite{Campagnola} and in biological membranes in the
presence of an electric potential.\cite{Peleg}

In this work, we study TPF and SHG in model molecular systems and
take advantage of the specific properties and complementarities of
these nonlinear processes. The main difference between them lies
in the coherent nature of the emitted signals and their relative
spectral features. Contrary to fluorescence, which is based on the
time-delayed relaxation of a radiative level, off-resonant SHG
occurs instantaneously upon excitation. Furthermore, the
fluorescence emission is affected by a Stokes shift, whereas SHG
occurs at half the incident wavelength and is phase-correlated
with the excitation field at the fundamental frequency. These
specific features allow one to distinguish them either spectrally
or temporally.\cite{VanHulst}

The main advantage of combining the two processes lies in their
respective anisotropy sensitivity to molecular orientation under
polarized excitation. In a centrosymmetric medium, SHG vanishes
while TPF is always present and can be anisotropic if the
molecular orientational distribution symmetry is
axial.\cite{Monson} Moreover, SHG appears in a polar medium, and
is intrinsically anisotropic. The difference between these two
effects is due to the distinct symmetry properties of the
tensorial susceptibilities that are involved, TPF and SHG being
described by even and odd order tensors respectively. The related
macroscopic susceptibilities reflect both molecular-scale and
macroscopic-scale arrangement properties, such as orientational
distribution. Measuring TPF and SHG anisotropies is therefore a
direct way to probe a given molecular organization in a molecular
ensemble that is either naturally ordered, or that has been
subject to the application of an external perturbation, such as an
electric field,\cite{Vorst,SingerKing} a coherent combination of
optical excitations,\cite{Brasselet} or a combination of
these.\cite{Dumont} Complex tensorial distributions can be
unravelled using such approaches.\cite{Bidault}

In order to fulfill the symmetry requirements discussed above, we
chose a model medium consisting of an amorphous polymer matrix
doped with fluorescent and/or dipolar nonlinear active
chromophores. Macroscopic centrosymmetry breaking of the medium is
induced  by electric field orientation of these chromophores,
using planar geometry suited to the two-photon microscopy
experimental setup. Applying an external electric field results in
a controllable external degree of freedom used to monitor the
molecular order symmetry. Furthermore, the reflection geometry of
the inverted microscopy configuration used in this work allows one
to measure both TPF and SHG signals without any dependence on the
coherence length of the nonlinear field propagation. The
experimental configuration presented here can be easily applied to
a variety of situations such as in biological membrane studies
where electric potentials might induce the orientation of a
molecular probe. Such studies can be potentially down-scaled to a
low number of molecules and even to single macromolecules, where
interactions at the molecular level or global conformational
changes can affect the collective emission anisotropy properties.

We present a detailed analysis of the orientational molecular
ordering in the sample using both TPF and SHG processes. We show
that the rotating-polarization analysis can bring additional
informations that might be hidden in ratiometric anisotropy
measurements. This analysis can be straightforwardly adapted to a
large variety of environments, and to dynamical
effects.\cite{SingerKing} Accounting for the instrumental features
of inverted microscopy, with high numerical aperture (N.A.)
objectives collecting light at wide angles of emission, we show
that a simple model allows one to retrieve information on
localized dipole orientations. As we demonstrate in this work, the
analysis of nonlinear anisotropies requires therefore particular
attention to the polarization state at the focus of a high N.A.
objective and the definition of a calibration procedure. In the
following sections, we discuss the optical properties of both TPF
and SHG processes in random and ordered media, taking into account
the influence of excitation transfer between chromophores.

\section*{II. Experimental section}
\subsection*{A. Sample Preparation}
The samples, studied at room temperature, contain active
chromophores, either
4-dicyanomethylene-2-methyl-6-(\emph{p}-(dimethylamino)styryl)-4H-pyran
(DCM) or 4-(N-ethyl-N-(2-hydroxyethyl))amino-4'-nitro-azobenzene
(Disperse Red~1, DR1). The DCM chromophore exhibits a high
fluorescence quantum yield and a non negligible molecular
quadratic polarizability $\beta$.\cite{DCM} The DR1 chromophore,
which has been extensively studied for its intrinsic high
quadratic nonlinear efficiency,\cite{DR1} is not fluorescent
because of quenching by fast photoisomerization. Both chromophores
exhibit a large permanent dipole moment which enables electric
field orientation.

The molecules are dispersed with a maximum concentration of $10\%$
by mass in a high $T_{\textrm{g}}$ (120$^{\circ}$C) polymer matrix
of polymethyl-methacrylate (PMMA), which is not cured after
deposition. The polymer films are then deposited by spincoating
either on a microscope slide or on an electrode
substrate.\cite{sample} In order to achieve electric field
alignment in the plane of the sample, we use a set of gold
transverse planar electrodes of 50 nm thickness, separated by a
gap of 10 $\mu$m, as represented in Figure~\ref{figure_montage}.
The electrodes are fabricated by photo-lithographic patterning of
UV sensitive photoresists on the gold layer, which is deposited by
sputtering on a microscope slide covered with a 100 nm PMMA
adhesion layer. The thickness of the active PMMA-dye layer is of
the order of 100 nm, in which the active molecules are expected to
orientate perpendicular to the electrodes in the sample plane.

\subsection*{B. Experimental Setup}
\begin{figure}[!ht]
\renewcommand{\captionlabeldelim}{.}
\centering
\includegraphics[width=13cm]{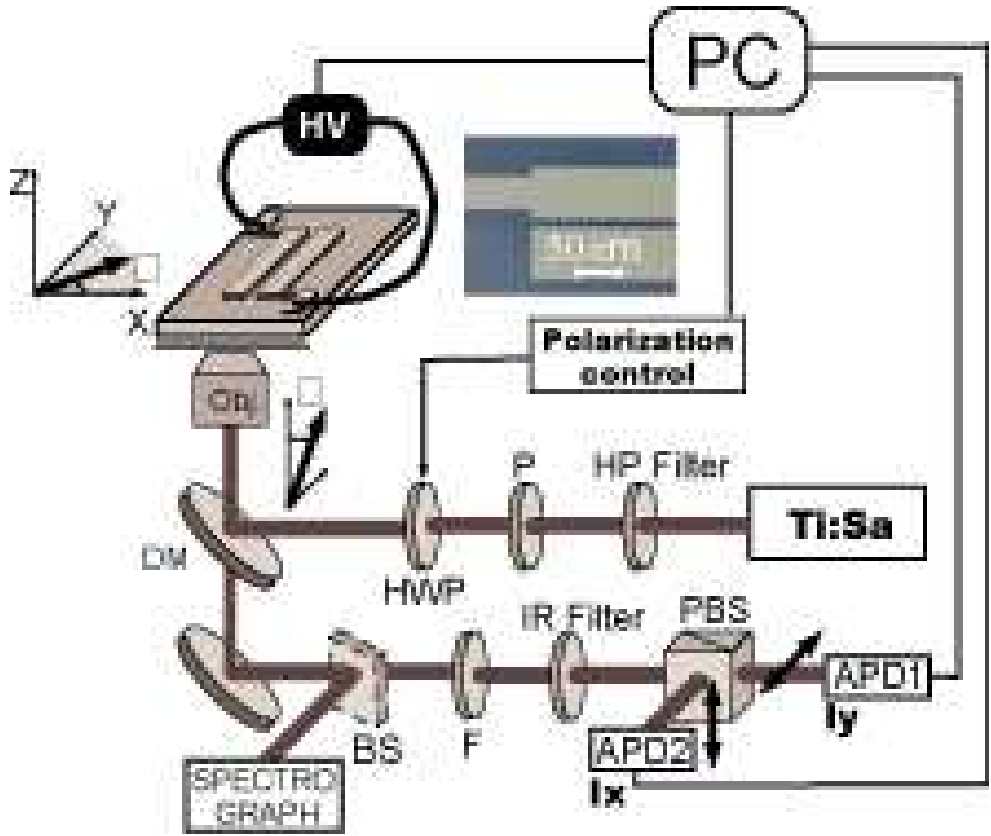}
\caption{Nonlinear microscopy setup: HP filter: high pass filter
which rejects visible light; P: polarizer; HWP: rotating half-wave
plate; Obj: microscope objective ($\times 60$, NA=1.4); DM:
dichroic mirror; F: interference filter selecting either the SHG
signal at 493 nm or the TPF signal at 580 nm; IR filter: filter
rejecting the residual incident (near IR) laser beam; PBS:
polarizing beamsplitter selecting $X$ and $Y$ polarization states
of the emission; APD: avalanche photodiodes. An additional
beamsplitter (BS) can be introduced for spectroscopic
measurements. During the polarization analysis, we remove this
beamsplitter in order to not perturb the polarization state of the
detected signal.} \label{figure_montage}
\end{figure}
The experimental setup used for nonlinear microscopy is described
in Figure~\ref{figure_montage}. The source for nonlinear
excitation is a mode-locked Ti:Sa laser which produces 120 fs
pulses at a fundamental wavelength of 987 nm with a 80 MHz
repetition rate. The laser beam is focused on the sample by a high
numerical aperture oil immersion objective, leading to a spatial
resolution of 400 nm. Typical incident energies range from 0.01 nJ
to 0.1 nJ per pulse. The TPF and SHG signals arising from the
sample are collected by the same objective, and then directed to a
polarizing beamsplitter and a set of two avalanche photodiodes
operating in the photon counting regime. We select either the SHG
signal or the TPF signal by appropriate interference filters. The
spectral distribution of the emitted light can be analyzed in
parallel, using a spectrograph coupled to a highly sensitive CCD
camera.

For both TPF and SHG, the polarization analysis consists of
rotating the incident polarization from 0$^{\circ}$ to
360$^{\circ}$ and recording the corresponding emissions on the two
perpendicular analyzed polarization directions. The geometry of
the system is schematically represented on
Figure~\ref{figure_montage}. The $Z$ axis is along the optical
axis, perpendicular to the sample. The $X$ and $Y$ axes, lying in
the sample plane, provide an analysis framework defining the
polarization directions detected by the two photodiodes. They also
coincide with  the $p$ and $s$ incident polarizations on the
dichroic mirror. The incident excitation polarization at the
fundamental frequency $\omega$ can be written at the focus point
as:\cite{frequence}
\begin{equation}
\vec{E}(\alpha,\delta,\gamma,\omega
t)=\frac{E}{\sqrt{1+(1-\gamma)^2}} \left[ \begin{array}{c}
\cos\alpha \cos(\omega t)\\
(1-\gamma) \sin\alpha \cos(\omega t + \delta)\\
\end{array} \right]
\end{equation}
where $E$ is the field amplitude and the rotating angle parameter
$\alpha$ defines the incident polarization angle in the ($X$,$Y$)
framework. In this model, we introduce two external parameters,
$\gamma$ and $\delta$, to account for polarization mixing effects
in the excitation/detection setup. The parameter $\gamma$
represents the difference of $p$ and $s$ reflectivity from the
dichroic mirror. The phase shift $\delta$ represents the
ellipticity appearing after reflection on the dichroic mirror with
an incidence angle of 45$^{\circ}$. The $X(p)$ and $Y(s)$
polarizations have been observed to be unaffected by any
ellipticity, allowing us to assume that ellipticity is only
introduced for intermediate $\alpha$ polarization angles. Note
that in this model, we neglect the longitudinal Gou\"y phase shift
factor $Z$ contribution of the incident field,\cite{Wolf} the
sample thickness being much lower than the coherence length of the
nonlinear process.\cite{gouy,MertzMoreaux} After passing through
the dichroic mirror, the light transmitted (reflected) by the
polarizing beamsplitter used for analysis is polarized in the $Y$
($X$) direction, corresponding to $I_Y$ ($I_X$) detected
intensities (see Figure~\ref{figure_montage}).

\section*{III. Results and Discussion}
\subsection*{A. Two-Photon Fluorescence Microscopy}
\begin{figure}[!ht]
\renewcommand{\captionlabeldelim}{.}
\begin{center}
\begin{minipage}[c]{0.5\textwidth}
\centering
\includegraphics[width=5cm]{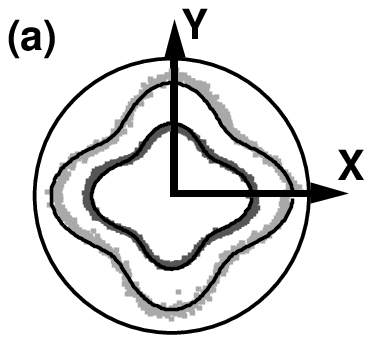}\\
\includegraphics[width=5cm]{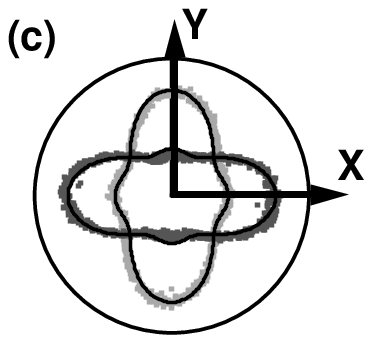}
\end{minipage}%
\begin{minipage}[c]{0.5\textwidth}
\centering
\includegraphics[width=5cm]{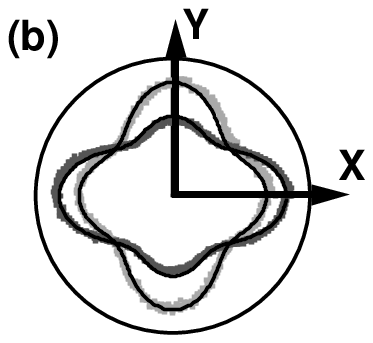}\\
\includegraphics[width=5cm]{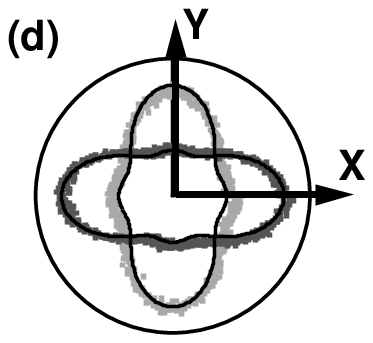}
\end{minipage}%
\caption{Fluorescence intensities $I_X$ (in dark grey) and $I_Y$
(in light grey) emitted by an isotropic assembly of DCM molecules,
shown in a polar plot as a function of the $\alpha$ angle varying
from $0^{\circ}$ to $360^{\circ}$. The integration time is 100 ms
per point. The four diagrams have been obtained for four different
concentrations in mass of DCM in PMMA: (a) 10$\%$ (b) 0.83$\%$ (c)
0.085$\%$ (d) 0.022$\%$. The continuous lines represent the fit
according to the expressions given in Appendix~A. The four graphs
have been normalized to the same intensity. The distinction
between the two intensity amplitudes $I_X$ and $I_Y$ in (a) is due
to a slight difference in detection efficiency in the two
polarization channels} \label{figure_dcm}
\end{center}
\end{figure}
In order to calibrate the polarization response of the nonlinear
microscopy setup and to account for the effect of the high
numerical aperture objective on the light polarization, we first
study the fluorescence emitted by an isotropic molecular
distribution. The sample used for this purpose is a guest-host
polymer matrix of PMMA embedded with the fluorescent dye DCM at
different concentrations. Figure~\ref{figure_dcm} shows the
experimental variations of $I_X$ and $I_Y$ fluorescence
intensities with respect to the polarization direction $\alpha$ of
the excitation beam. In order to explain the observed features,
several experimental considerations have to be accounted for, in
addition to the incident field polarization parameters detailed
previously.

First, the high numerical aperture of the objective affects the
polarization radiated by the dye molecules.\cite{Wolf} This effect
can be modelled using a calculation similar to Ref.~\cite{Axelrod}
relative to one-photon fluorescence. Indeed, TPF emission is still
occurring from a one-photon allowed transition, independently of
the excitation pathway. We define the direction of the emission
dipole of the rod-like DCM molecule by a set of two angles
$(\theta,\phi)=\Omega$. The fluorescence intensities $J_X$ and
$J_Y$ emitted by this single dipole
$\vec{\mu}(\Omega)=(\mu_X(\Omega),\mu_Y(\Omega),\mu_Z(\Omega))$
and measured by the two detectors can be written as expressed in
Appendix~A:
\begin{eqnarray}
J_X(\Omega)=K_1\, \mu_X^2(\Omega)+K_2\, \mu_Y^2(\Omega)+K_3\, \mu_Z^2(\Omega)\nonumber\\
J_Y(\Omega)=K_2\, \mu_X^2(\Omega)+K_1\, \mu_Y^2(\Omega)+K_3\,
\mu_Z^2(\Omega)
\end{eqnarray}
where $K_1$=2.945, $K_2$=0.069 and $K_3$=1.016 in the case of a
NA=1.4 objective such as the one used in the present work. $K_1$,
$K_2$ and $K_3$ represent the mixing of polarization components in
the emission, as a consequence of the collection of light at very
wide angle.

Second, the two-photon nature of the excitation process has to be
taken into account. Contrary to a 1-photon excitation which
depends on the square of the incident field $\vec{E}$, the
2-photon excitation probability of a dipole $\vec{\mu}$ is
proportional to $|\vec{\mu}.\vec{E}|^4$. Therefore, the
polarization state of the incident IR beam plays a crucial role.
We will furthermore suppose that the emission and excitation
dipoles are parallel for DCM molecules.\cite{angle}

Third, we assume a normalized molecular orientational distribution
function $f(\Omega )$ which is set equal to $1/4\pi$ in the case
of an isotropic molecular ensemble in the polymer matrix. Using
the previous assumptions, the respective fluorescence intensities
in the $X$ and $Y$ analysis directions can be expressed~as:
\begin{equation}
I_{I=X,Y}^{\textrm{TPF\,(direct)}}=\int \overline{\arrowvert
\vec{\mu}(\Omega).\vec{E} \arrowvert^4} J_I(\Omega) f(\Omega)
\textrm{d}\Omega
\end{equation}
after orientational averaging with $\textrm{d}\Omega=\sin\theta\,
\textrm{d}\theta \,\textrm{d}\phi$ and subsequent temporal average
represented by the $\overline{(...)}$ notation.

\begin{figure}[!ht]
\renewcommand{\captionlabeldelim}{.}
\begin{center}
\mbox{\includegraphics[width=10cm]{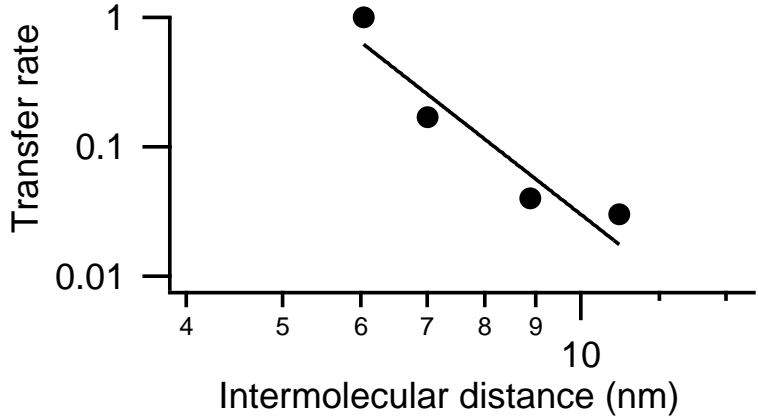}} \caption{Log-log
plot of the experimental transfer rate versus estimated
intermolecular distance, as the DCM concentration in the polymer
matrix is varied. A $1/r^6$ power dependance is also plotted as a
guideline.} \label{figure_transfer}
\end{center}
\end{figure}

Finally, Figure~\ref{figure_dcm} shows that the polarization
response depends strongly on the molecular concentration, which is
the signature of a possible excitation transfer between
fluorescent molecules. The effect of such transfer in highly
concentrated media is to decouple the excitation step from the
emission one,\cite{Transfer} resulting in a depolarization
process, which leads to the same signal in the two analyzing
channels. This is indeed what is observed in
Figure~\ref{figure_dcm}a. The excitation transfer efficiency
between neighboring-molecules depends on the relative orientation
factor $\kappa^2$ between two neighboring  chromophores, the
``donor" $\vec{\mu}_1(\theta_1,\phi_1)$) and the ``acceptor"
$\vec{\mu}_2(\theta_2,\phi_2,\rho,\xi,\theta_1,\phi_1)$):\cite{Clegg}
\begin{equation}
\kappa^2(\rho,\theta_2,\phi_2)=\left( 2
\cos\rho.\sin\theta_2.\cos\phi_2+
\sin\rho.\sin\theta_2.\sin\phi_2\right)^2
\end{equation}
In this equation, $(\theta_1,\phi_1)$=$\Omega_1$ defines the
orientation of the donor $\vec{\mu}_1$, $(\rho,\xi)$=$\Omega$
defines the orientation of the vector $\vec{u}_{12}$ connecting
the two chromophores in the ($\theta_1,\phi_1$) framework and
$(\theta_2,\phi_2)$=$\Omega_2$ defines the orientation of the
acceptor $\vec{\mu}_2$ in the ($\rho,\xi$) framework. The
expression of the vector
$\vec{\mu}_2(\theta_2,\phi_2,\rho,\xi,\theta_1,\phi_1)$ in the
macroscopic framework $(X,Y,Z)$ is given in Appendix~B. The
fluorescence intensity after energy transfer can then be
written~as:
\begin{equation}
I_{I=X,Y}^{\textrm{TPF\,(transfer)}}=\iiint \overline{\arrowvert
\vec{\mu_1(\Omega_1)}.\vec{E} \arrowvert^4}
\kappa^2(\rho,\theta_2,\phi_2)J_I(\Omega_2, \Omega_1, \Omega)
 f(\Omega_1) f(\Omega_2)\textrm{d}\Omega\textrm{d}\Omega_1 \textrm{d}\Omega_2\,\,\,
\end{equation}
where $J_I(\Omega_2, \Omega_1, \Omega)$ is given in Appendix~B.
The total fluorescence intensity can then be written~as:
\begin{equation}
I^{\textrm{TPF}}=I^{\textrm{TPF\,(direct)}}+\mathcal{T} \,
I^{\textrm{TPF\,(transfer)}}
\end{equation}
where $\mathcal{T}$ is a fitting parameter which quantifies the
transfer rate. As seen in Figure~\ref{figure_dcm}, this model is
in good agreement with the data, using the experimental factors
$\delta$=1.15 rad for the field phase shift responsible for the
incident ellipticity and $\gamma$=0.02 for the dichroism. These
parameters have been measured separately by ellipsometry at our
fundamental wavelength. The effect of ellipticity is to give to
the polarization response a cross shape instead of the expected
uniform circle in the case of randomly oriented isolated
molecules. Figure~\ref{figure_dcm}a corresponds to a high
concentration sample resulting in a complete depolarization of the
emission for which a transfer rate $\mathcal{T}$=1 allows one to
fit the data. Due to the intermolecular long-range coupling
between the two dipoles, the energy transfer rate $\mathcal{T}$
depends mainly on the inter-molecular distance $r$ in the polymer
and is proportional to $1/r^6$. $r$ was varied by decreasing the
DCM concentration in mass, namely from 10$\%$ to 0.83$\%$ and
0.085$\%$, down to 0.022$\%$. In each case, an average
intermolecular distance can be estimated, giving values of 6.0 nm,
7.0 nm, 8.9 nm and 10.9 nm respectively. The effect of a
decreasing concentration appears clearly in
Figure~\ref{figure_dcm}, with the lateral lobes of $I_X$ and $I_Y$
disappearing progressively in the $Y$ and $X$ directions
respectively. Note that for the concentration corresponding to
Figure~\ref{figure_dcm}d, the excitation transfer is negligible,
and each detector detects mainly its preferential polarization
direction as expected for an incoherent optical process. The
energy transfer rates obtained from the fits of
Figure~\ref{figure_dcm} are respectively $1$, $0.15$, $0.03$ and
$0.02$. These values are represented on
Figure~\ref{figure_transfer}. The dependance with the
intermolecular distance $r$ is compared with a $1/r^6$ power law,
showing a qualitative good agreement.

\subsection*{B. Second Harmonic Generation Microscopy}
\begin{figure}[!ht]
\renewcommand{\captionlabeldelim}{.}
\begin{center}
\begin{minipage}[c]{0.5\textwidth}
\centering
\includegraphics[height=2.5cm]{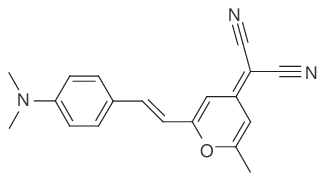}\\
\includegraphics[width=5cm]{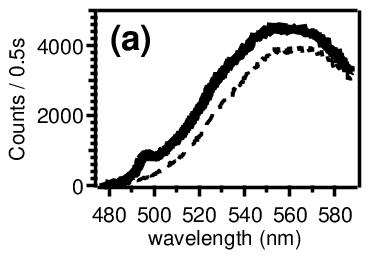}\\
\includegraphics[width=5cm]{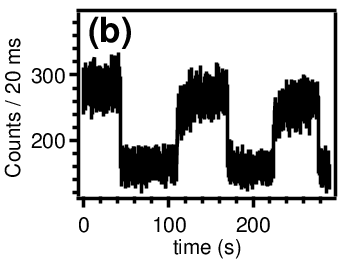}
\end{minipage}%
\begin{minipage}[c]{0.5\textwidth}
\centering
\includegraphics[height=2.5cm]{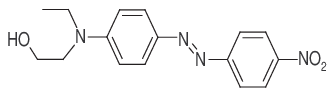}\\
\includegraphics[width=5cm]{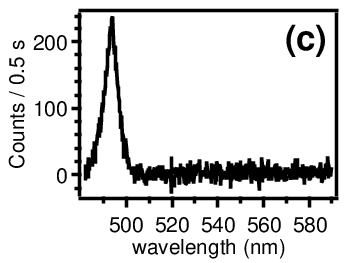}\\
\includegraphics[width=5cm]{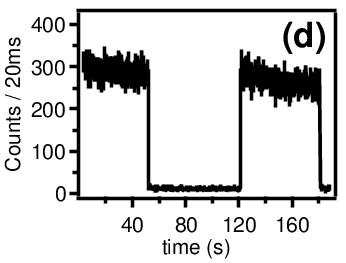}
\end{minipage}%
\caption{(a) Two-photon emission from a polymer film doped with
DCM. A SHG peak appears next to the fluorescence band when a high
voltage is applied to the sample (dashed line: 0~V; continuous
line: 300~V). (b) SHG modulation from DCM molecules under an
ON-OFF poling field. The high background is due to the DCM
residual TPF, which lies within the filter bandpass. Note that it
would be possible to avoid such a signal by working off resonance
for the SHG. (c) Two-photon emission from a polymer film doped
with DR1, showing only SHG. The width of the SHG peak is limited
by the incoming IR femtosecond laser linewidth. (d) SHG modulation
for DR1 molecule. In both cases, the incident IR laser beam energy
is about 0.1 nJ per pulse.} \label{figure_molecule}
\end{center}
\end{figure}
The same setup can also be used to study the influence of a
non-centro\-symme\-tric contribution of the molecular distribution
on the optical response upon two-photon excitation. In this case,
a SHG signal arises from the polar orientation of nonlinear active
molecules. For this purpose, the DCM doped polymer matrix is
spincoated on a set of two transverse electrodes, as described in
the first part of this paper. A high voltage of about 300 V,
corresponding to a poling field of $3\times10^7$ V.m$^{-1}$, is
applied by means of electrical contacts on the electrodes. As the
DCM molecules have a non negligible permanent dipole moment of
about 10 Debye,\cite{DCM} they will have a tendency to lean
towards the electric field direction, thus creating a
non-centrosymmetric distribution. Moreover, due to the nonlinear
susceptibility $\beta$ of DCM,\cite{DCM} the molecular orientation
can be directly monitored by the onset of a SHG signal in addition
to fluorescence as reported in the previous section. In the
present configuration, the SHG signal is detected in a reflection
configuration. Since the thickness of the molecular layer is much
smaller than the coherence length for phase-matching, the
molecules can be simply considered as a coherent ensemble of
nonlinear radiating dipoles. The SHG signal therefore appears with
the same intensity either in the transmitted or in the reflected
direction. As shown on Figure~\ref{figure_molecule}a, the overall
two-photon emission spectrum exhibits a narrow SHG peak at
$2\omega$, emerging in the tail of the broadband fluorescence when
the electric field is applied. The SHG and TPF signals can be
easily selected by introducing appropriate optical filters.
Figure~\ref{figure_molecule}b shows the temporal evolution of SHG
upon the ON-OFF application of the electric field: since the SHG
signal disappears when the electric field is turned off, on can
assume that the molecules still have a high degree of mobility in
the polymer matrix in spite of its high $T_{\textrm{g}}$. This
mobility is probably a result of the broad size distribution of
cavities in the polymer with a tail in the large size limit.

\begin{figure}[!ht]
\renewcommand{\captionlabeldelim}{.}
\begin{center}
\begin{minipage}[c]{0.3\textwidth}
\centering
\includegraphics[width=4cm]{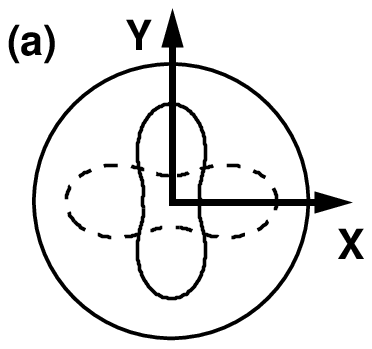}
\end{minipage}%
\hspace{0.04\linewidth}%
\begin{minipage}[c]{0.3\textwidth}
\centering
\includegraphics[width=4cm]{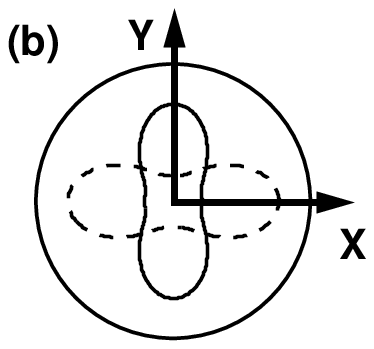}
\end{minipage}%
\hspace{0.04\linewidth}%
\begin{minipage}[c]{0.3\textwidth}
\centering
\includegraphics[width=4cm]{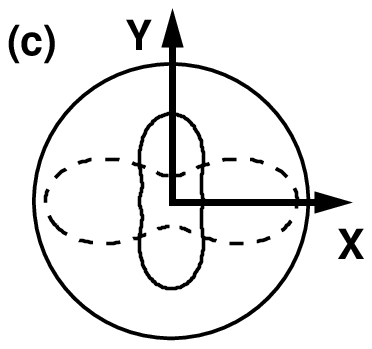}
\end{minipage}%
\caption{Influence of the poling field $\vec{E}_0$, aligned along
the $X$ axis, on polar polarization plots of the calculated TPF
intensities $I_X$ (dashed line)  and $I_Y$ (continuous line)
emitted by an assembly of fluorescent molecules. Energy transfer
effects are not taken into account in these calculations. All
graphs have the same scale (a) Fluorescence diagram for an
isotropic molecular distribution. (b) Fluorescence diagram with a
field of $3\times10^7$ V.m$^{-1}$, such as used in our experiment.
There is no clear difference with the first diagram. (c) The
effect of the applied field becomes visible for $3\times10^8$
V.m$^{-1}$, corresponding to a voltage 10 times higher than the
experimentally used values.} \label{figure_dcm_th}
\end{center}
\end{figure}

When spectrally selecting the TPF signal only, the polarization
response of the sample does not undergo significant changes in the
presence of the static electric field. The fluorescence
sensitivity to molecular orientation can be explored by
calculating similar polarization patterns as those observed in
Figure~\ref{figure_dcm}, accounting for the molecular
orientational distribution induced by electric field orientation.
The calculated fluorescence patterns, shown on
Figure~\ref{figure_dcm_th}, have been computed considering a
permanent poling field $\vec{E}_0$ and using the Boltzmann
distribution function:
\begin{equation}
f(\Omega)=\frac{\textrm{e}^{(\vec{\mu}(\Omega).\vec{E}_0)/k_BT)}}{\int
\textrm{e}^{(\vec{\mu}(\Omega).\vec{E}_0)/k_BT)} \textrm{d}\Omega}
\label{bolztmann}
\end{equation}
where $k_B$ is the Boltzmann's constant and $T$ the poling
temperature. As seen from Figure~\ref{figure_dcm_th}b in our
experimental conditions, i.e. with an electric field of
$3\times10^7$ V.m$^{-1}$ and at room temperature, the fluorescence
patterns exhibit only slight differences compared to an isotropic
molecular distribution. This result is not surprising since the
TPF intensity is proportional to the number $N$ of active
molecules per unit volume, whereas SHG intensity is proportional
to $N^2$ as a result of the coherence of this radiating process.
Moreover, the weak effect of the electric field on the
fluorescence anisotropy is also due to the broad statistical
orientational Boltzmann distribution at room temperature.
Figure~\ref{figure_dcm_th}c shows that orientational effects
should be strongly increased with higher electric fields, which
are however difficult to reach due to electric breakdown.

The SHG signal is therefore a better suited probe of the
field-induced anisotropic molecular distribution.
\begin{figure}[!ht]
\renewcommand{\captionlabeldelim}{.}
\begin{center}
\begin{minipage}[c]{0.3\textwidth}
\centering
\includegraphics[width=4cm]{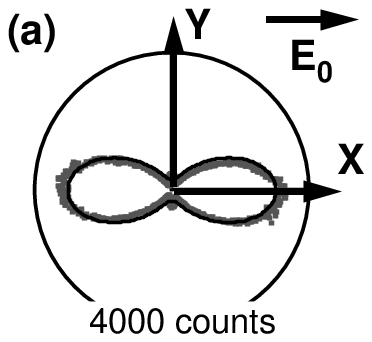}\\
\includegraphics[width=4cm]{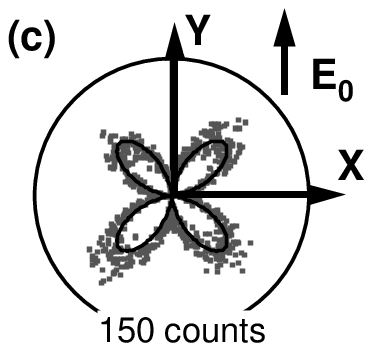}
\end{minipage}%
\hspace{0.04\linewidth}%
\begin{minipage}[c]{0.3\textwidth}
\centering
\includegraphics[width=4cm]{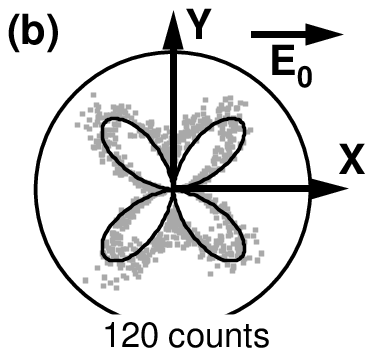}\\
\includegraphics[width=4cm]{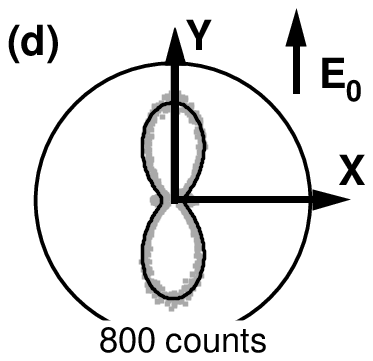}
\end{minipage}%
\hspace{0.04\linewidth}%
\begin{minipage}[c]{0.3\textwidth}
\centering
\includegraphics[width=4cm]{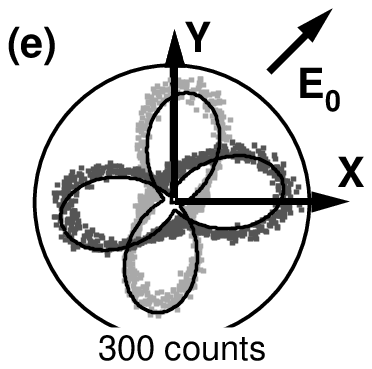}
\end{minipage}%
\caption{Polar plots showing the SHG signals (integration time:
100~ms) $I_X$ (in dark grey) and $I_Y$ (in light grey) emitted by
an assembly of DR1 molecules poled under an electric field of
$3\times10^7$ V.m$^{-1}$ as a function of the incident
polarization angle varying from $0^{\circ}$ to $360^{\circ}$. The
different diagrams have been obtained for different orientations
of the electric field, as indicated on the Figures. (a) and (b):
$I_X$ and $I_Y$ for $\vec{E}_0$ along $X$, (c) and (d): $I_X$ and
$I_Y$ for $\vec{E}_0$ along $Y$, (e): $I_X$ and $I_Y$ for
$\vec{E}_0$ at $45^{\circ}$ of $X$ and $Y$. The continuous lines
represent the fit according to the expressions given in Appendix
C. The situations where $\vec{E}_0$ is along the $X$ or $Y$ axis
are symmetric, whereas a rotation of 45$^{\circ}$ mixes the
nonlinear polarization responses together.} \label{figure_dr1}
\end{center}
\end{figure}
In order to retrieve SHG information independently from
fluorescence properties, we choose to study more specifically
samples doped with the DR1 molecules, so as to prevent from
possibly misleading combinations of SHG and TPF signals. This
situation is akin to using DCM molecules with non-resonant
excitation, where TPF would then be inefficient. DR1 has a
non-negligible dipole moment of 8.7 Debye,\cite{DR1} and a
molecular hyperpolarizability $\beta$=289$\times10^{-40}$
m$^4$V$^{-1}$ at zero frequency, which amounts to twice that of
DCM.\cite{DCM} Figures~\ref{figure_molecule}c and
\ref{figure_molecule}d show the effect of electric field po\-ling
of DR1 using the same electrode system . The same polarization
analysis is applied to investigate the SHG responses, and
Figure~\ref{figure_dr1} shows the evolution of the SHG signal with
respect to the rotation of the incident polarization $\alpha$.

In order to fit the data, we developed a model elaborating on Ref.
\cite{Vorst,Singer}, based on the macroscopic second order
susceptibility $\chi^{(2)}(2\omega ; \omega, \omega)$, which is
related to the molecular hy\-per\-po\-la\-ri\-za\-bi\-li\-ty
$\beta(2\omega ; \omega, \omega)$ tensor and the molecular
distribution function $f(\Omega)$. Earlier models are extended
herein so as to account for specific features pertaining to the
two-photon microscopy setup, as detailed in Appendix~C. The
rod-like DR1 molecule is composed of a conjugated $\pi$ system
with an electron acceptor group at one end and a donor group at
the other end. The nonlinear susceptibility $\beta$ has therefore
only one non-zero component $\beta_{zzz}$, where $z$ defines  the
molecular axis direction. We denote $\Omega=(\theta ,\phi )$ the
Euler angles defining $z$ in the macroscopic framework. With $N$
molecules per unit volume, the $\chi^{(2)}$ susceptibility tensor
components can be expressed as:
\begin{equation}
\chi_{\scriptscriptstyle{IJK}}^{(2)}=N \beta_{zzz} \int \cos(z,I)
\cos(z,J) \cos(z,K)f(\Omega)\textrm{d}\Omega
\end{equation}
where the indices $I,J,K =X,Y,Z$ are the coordinates in the
macroscopic framework and $f(\Omega)$ is the molecular
orientational distribution function given by
Eq.~(\ref{bolztmann}). The $\cos(z,I)$ functions are the
$(\Omega)$ angle dependent projections of the $z$ axis on the $I$
axis. For symmetry reasons, a non-zero value for $\chi^{(2)}$
requires a macroscopic polar order, which is imposed here by the
static electric field $\vec{E}_0$ applied in the plane of the
sample. From the $\chi_{\scriptscriptstyle{IJK}}^{(2)}$ tensor
coefficients, we can infer the induced nonlinear polarization
$\vec{P}^{(2)}(2\omega)$ defined~as:
\begin{equation}
 P_I^{(2)}(2\omega)=\sum_{J,K}
\chi_{\scriptscriptstyle{IJK}}^{(2)}\,E_J(\omega)\,E_K(\omega)
\end{equation} Note that this expression of $P_I^{(2)}(2\omega)$
does not take into account the local field factors
$f_I^{2\omega}$, $f_J^{\omega}$ and $f_K^{\omega}$, which is
reasonable in the present context of relatively low molecular
concentration and weak poling strength.\cite{Stegeman} As detailed
in Appendix~C, this polarization allows us to compute the
$2\omega$ radiated field. The correction factors accounting for
the collection of light at wide angles are estimated with a model
similar to the one used for TPF. Since SHG is a coherent process
arising from the induced dipole radiation, the collected field can
be written~as:
\begin{equation}
\mathcal{E}^{SHG}_I(2\omega)=b_{IXX} E_X^2 + b_{IYY} E_Y^2 +
b_{IXY} E_X E_Y
\end{equation}
where $b_{IJK}$ are constant coefficients defined in Appendix~C.
The resulting intensity is proportional to $|\vec{\mathcal{E}}^{
SHG}(2\omega)|^2$, and the SHG intensities $I_I^{\textrm{SHG}}$
detected by the photodiodes after temporal averaging can be
finally expressed as:
\begin{eqnarray}
I_{I=X,Y}^{\textrm{SHG}}&=&b_{IXX}^2\,   \overline{E_X^4}   +
b_{IYY}^2\, \overline{E_Y^4}
 + (2\,b_{IXX}\,b_{IYY} + b_{IXY}^2)\, \overline{E_X^2E_Y^2}\nonumber\\
 &+& 2\,b_{IXX}\,b_{IXY}\,  \overline{E_X^3E_Y } +
2\,b_{IYY}\,b_{IXY}\,  \overline{E_XE_Y^3}
\end{eqnarray}
Figure~\ref{figure_dr1} represents the experimental data and the
theoretical fits for several orientations of the poling field
$\vec{E}_0$ relative to the ($X,Y$) axes, showing a good
agreement. These fits also take into account the ellipticity
parameter $\delta=1.15$ rad and the $\gamma$ factor of 0.02
determined previously. Contrary to fluorescence, the low
intermolecular distance in the polymer matrix is seen to not
affect the polarization response of SHG. These results show that a
thorough SHG polarization analysis is able to account for a slight
change in molecular organization. In particular, the symmetry
features of SHG polarization polar plots shown in
Figure~\ref{figure_dr1} are consistent with the rotation of the
electric field in the ($X,Y$) plane. Such polarization signatures
can be exploited to identify specific molecular orientation
directions in the sample plane. In the present case, the $2\omega$
optical response is spatially uniform over a raster scan of 5
$\mu$m$\times$5 $\mu$m (data not shown). This provides clear
evidence of the good homogeneity of the field orientation between
the two planar electrodes. Such studies are being currently
extended to the mapping of molecular polar-orientation in
electro-optic devices based on complex electrodes
designs,\cite{Donval} or to lower-scale changes in molecular
orientation patterns such as those observed in Langmuir-Blodgett
films.\cite{Bozhevolnyi}

\section*{IV. Conclusion}
Polarization analysis of the emission response from an assembly of
molecules under two-photon excitation contains significant
information on their orientational distribution. Focusing on
either TPF or SHG furthermore allows one to study either the
incoherent, axial even-order processes, or the coherent polar
odd-order processes. Combination of these two techniques provides
a complete set of informations regarding possible excitation
transfer between molecules or orientation of a collection of
molecules with very accute sensitivity to the onset of
non-centrosymmetric orientational patterns.

The approach presented in this paper can be applied to a broad
variety of molecular media, where unknown orientational
distribution functions are expressed using spherical harmonics
which can be singled-out via their symmetry
properties.\cite{Bidault} By taking advantage of the high optical
resolution provided by two-photon microscopy, interesting
perspectives can be expected in the study of structures involving
sub-microscopic scale effects such as in
nanocrystals\cite{Prasad,Treussart} and nanostructured media
supporting local field enhancements.\cite{Anceau}

\subsection*{Appendix A: Polarization response of TPF
 in a high NA microscope setup for wide angle fluorescence collection}
The following model is developed for comparison between
polarization responses in two perpendicular polarization
directions (see Figure~\ref{figure_montage}). In the subsequent
derivations, we omit all efficiency parameters that may appear in
the intensity expressions, since such parameters are the same for
each channel of polarization direction.
\begin{figure}[!ht]
\renewcommand{\captionlabeldelim}{.}
\begin{center}
\includegraphics[width=6cm]{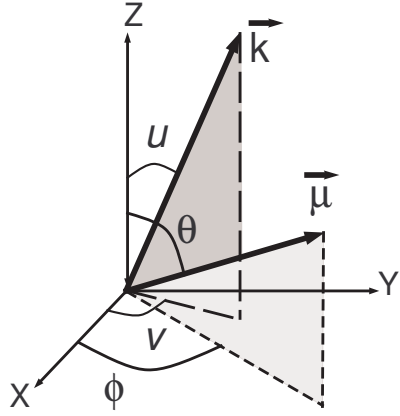}
\caption{Notations introduced to account for the wide angle
collection of the emitted light by a dipole, located at the origin
of the $(X,Y,Z)$ framework. The direction of the dipole
$\vec{\mu}$ is defined by the angles $(\theta,\phi)$. The
direction of the wave vector $\vec{k}$ corresponding to the
radiated field, is defined by the angles $(u,v)$.}
\label{figure_dessin1}
\end{center}
\end{figure}

In order to express the intensity of fluorescence emitted by a
single dipole set at the focal point of the microscope objective,
we consider the radiation diagram emitted by this dipole
$\vec{\mu}$, which orientation is defined by the angles
$(\theta,\phi)$, as indicated on Figure~\ref{figure_dessin1}. The
far field radiated in the direction of the wave vector
$\vec{k}(u,v)$ is:
\begin{equation}
\vec{\mathcal{E}}^{\textrm{ TPF (radiated)}}\propto\vec{k} \wedge
(\vec{\mu} \wedge \vec{k})
\end{equation}
were the proportionality factor contains constant radiation field
terms that are not used in the following calculations. As the
orientation of the wave vector $\vec{k}$ is only defined by the
angles ($u,v$), the vector $\vec{\mathcal{E}}^{\textrm{ TPF
(radiated)}}$ can be expressed as:
\begin{equation}
\vec{\mathcal{E}}^{\textrm{ TPF (radiated)}}(u,v, \theta, \phi) =
\mu_X(\theta, \phi) \, \vec{U_1}(u,v) + \mu_Y(\theta, \phi) \,
\vec{U_2}(u,v) + \mu_Z(\theta, \phi) \vec{U_3}(u,v)
\end{equation}
where $\vec{U_1}$, $\vec{U_2}$, $\vec{U_3}$ are unit vectors
depending only on the parameters $(u,v)$.
The field transmitted by the objective can then be expressed as:
\begin{equation}
\vec{\mathcal{E}}^{\textrm{ TPF}}=
[\mathcal{R}]\,\vec{\mathcal{E}}^{\textrm{ TPF(radiated)}}
\label{objective}
\end{equation}
 where
$[\mathcal{R}]$ represents the rotation matrix simulating the
infinity-corrected objective. $[\mathcal{R}]$ is therefore the
product of three successive rotations\cite{Axelrod} (rotation of
$-v$ around $Z$, rotation of $-u$ around $Y$ and rotation of $v$
around $Z$) so as to convert any input incidence on the objective
into an output ray parallel to the optical axis $Z$, namely:
\begin{equation}
[\mathcal{R}]= \left[
\begin{array}{c c c}
\cos u\, \cos^2 v + \sin^2 v & \cos v\, \sin v (\cos u
-1) & -\sin u\, \cos v\\
\cos v\, \sin v (\cos u -1) & \cos u\, \sin^2 v+
\cos^2 v & -\sin u\, \sin v\\
\sin u\, \cos v & \sin u\, \sin v & \cos u
 \end{array} \right]
\end{equation}
  The vector
$\vec{\mathcal{E}}^{\textrm{ TPF}}(u,v, \theta, \phi)$ of the
trasmitted field can then be expressed as:
\begin{eqnarray}
\mathcal{E}^{\textrm{TPF}}_X(u,v, \theta, \phi)=f_X(u,v)\,
\mu_X(\theta, \phi)+f_Y(u,v)\, \mu_Y(\theta, \phi)+f_Z(u,v)\,
\mu_Z(\theta, \phi)\nonumber\\
\mathcal{E}^{\textrm{TPF}}_Y(u,v, \theta, \phi)=g_X(u,v)\,
\mu_X(\theta, \phi)+g_Y(u,v)\, \mu_Y(\theta, \phi)+g_Z(u,v)\,
\mu_Z(\theta, \phi)
\end{eqnarray}
where $f_X$, $f_Y$, $f_Z$, $g_X$, $g_Y$ and $g_Z$  are functions
of the $(u,v)$ parameters.

Since fluorescence light is emitted incoherently, the intensities
detected by the photodetectors and coming from the single dipole
$\vec{\mu}(\theta,\phi)$, are computed after integration of the
square of each $\vec{\mathcal{E}}^{\textrm{ TPF}}$ component, over
all the angles $(u,v)$ within the half-aperture angle
$\theta_{\textrm{obj}}$ of the objective, hence giving the
detection probability:
\begin{equation}
J_{I=X,Y}(\theta, \phi)=\int_0^{2\pi}
\int_0^{\theta_{\textrm{obj}}}
\big(\mathcal{E}^{\textrm{TPF}}_I(u,v,\theta, \phi)\big)^2
\sin{u}\, \textrm{d}u \,\textrm{d}v
\end{equation}
In the present work, the oil-immersion ($n$=1.5) objective has a
numerical aperture NA=1.4. The half-aperture angle
$\theta_{\textrm{obj}}$ is therefore equal to $1.204$ rad
(69$^{\circ}$). After integration, the previous expression reduces
to:
\begin{eqnarray}
J_X(\theta, \phi)=K_1\, \mu_X^2(\theta, \phi)+K_2\,
\mu_Y^2(\theta, \phi)+K_3\, \mu_Z^2(\theta, \phi)\nonumber\\
J_Y(\theta, \phi)=K_2\, \mu_X^2(\theta, \phi)+K_1\,
\mu_Y^2(\theta, \phi)+K_3\, \mu_Z^2(\theta, \phi)
\end{eqnarray}
with:
\begin{eqnarray}
K_1=\int_0^{2\pi} \int_0^{\theta_{\textrm{obj}}}f_X^2(u,v)
\sin{u}\, \textrm{d}u \,\textrm{d}v=\int_0^{2\pi}
\int_0^{\theta_{\textrm{obj}}}g_Y^2(u,v)\sin{u}\, \textrm{d}u\,
\textrm{d}v=2.945\nonumber\\
K_2=\int_0^{2\pi} \int_0^{\theta_{\textrm{obj}}}f_Y^2(u,v)
\sin{u}\, \textrm{d}u\, \textrm{d}v=\int_0^{2\pi}
\int_0^{\theta_{\textrm{obj}}}g_X^2(u,v) \sin{u}\, \textrm{d}u\,
\textrm{d}v=0.069\nonumber\\
K_3=\int_0^{2\pi} \int_0^{\theta_{\textrm{obj}}}f_Z^2(u,v)
\sin{u}\, \textrm{d}u\, \textrm{d}v=\int_0^{2\pi}
\int_0^{\theta_{\textrm{obj}}}g_Z^2(u,v) \sin{u}\, \textrm{d}u\,
\textrm{d}v=1.016\,\,
\end{eqnarray}

Moreover, the two photon excitation probability, which is
proportional to $|\vec{\mu}(\Omega).\vec{E}|^4$, can be easily
evaluated. Since fluorescence is an incoherent process, we
directly integrate the dipole response, product of the excitation
 and detection probability, over all dipole orientations.
Assuming that the absorption and emission dipoles of each
chromophore are parallel, the detected intensities can then be
expressed as:
\begin{equation}
I_{I=X,Y}^{\textrm{TPF}}= \int \overline{\arrowvert
\vec{\mu}(\Omega).\vec{E}
 \arrowvert^4} J_I(\Omega) f(\Omega) \textrm{d}\Omega
\end{equation}
Developing $\overline{|\vec{\mu}(\Omega).\vec{E}|^4}$ in the
previous expressions, the TPF intensities, depending only on
parameters $(\alpha,\delta,\gamma)$, finally reduce to:
\begin{eqnarray}
I_{I=X,Y}^{\textrm{TPF}}(\alpha,\delta,\gamma)&=&\sum_{J,K,L,M=X,Y}f_{IJKLM}\,
\overline{E_J E_K E_L E_M}\nonumber\\
&=&f_{IXXXX} \overline{E_X^4}+ f_{IYYYY}\overline{E_Y^4} + 6
f_{IXXYY}\overline{E_X^2 E_Y^2}\nonumber\\
&+& 4
f_{IXXXY}\overline{E_X^3E_Y}+ 4f_{IXYYY}\overline{E_X E_Y^3}
\end{eqnarray}
where the $f_{IJKLM}$ coefficients are defined by:
\begin{equation}
f_{IJKLM}=\int J_I(\Omega)\, \mu_J(\Omega)\, \mu_K(\Omega)\,
\mu_L(\Omega)\, \mu_M(\Omega)\, f(\Omega)\, \textrm{d}\Omega
\end{equation}

\subsection*{Appendix B: Expression of the acceptor  coordinates
$\vec{\mu}_2$ in the macroscopic framework}
\begin{figure}[!ht]
\renewcommand{\captionlabeldelim}{.}
\centering
\includegraphics[width=8cm]{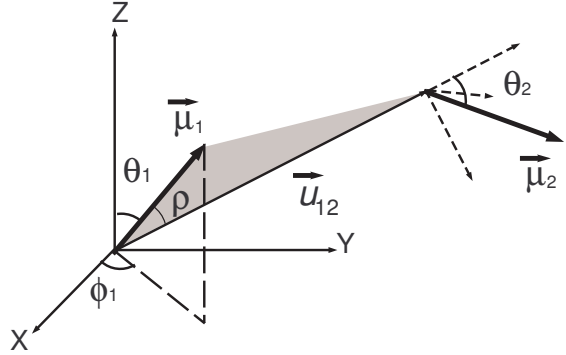}
\caption{Position and orientation of the acceptor $\vec{\mu}_2$
relative to the position and orientation of the donor
$\vec{\mu}_1$. The orientation of the donor $\vec{\mu}_1$ is
defined by the ($\theta_1,\phi_1$) angles. The orientation of the
unitary vector $\vec{u}_{12}$ along the line connecting the two
chromophores is defined by the ($\rho,\xi$) angles in the
($\theta_1,\phi_1$) framework. The orientation of the acceptor
$\vec{\mu}_2$ in the ($\rho,\xi$) framework is defined by the
$(\theta_2,\phi_2)$ angles in the local framework of molecule 2
(shown in dotted lines).} \label{figure_dessin2}
\end{figure}
As indicated in Figure~\ref{figure_dessin2}, ($\theta_1,\phi_1$)
defines the orientation of the donor dipole $\vec{\mu}_1$ and
($\rho,\xi$) defines the orientation of the vector $\vec{u}_{12}$
connecting the two chromophores in the ($\theta_1,\phi_1$)
framework. The orientation of the acceptor dipole $\vec{\mu}_2$ is
defined by the angles $(\theta_2,\phi_2)$ in the ($\rho,\xi$)
local framework of molecule 2. The acceptor coordinates in the
macroscopic framework depend consequently on the angles
$(\theta_2,\phi_2,\rho,\xi,\theta_1,\phi_1)$ and can be written
as:
\begin{eqnarray}
\vec{\mu}_2(\theta_2,\phi_2,\rho,\xi,\theta_1,\phi_1)=
[\mathcal{M}]_{(\theta_1,\phi_1)\to(X,Y,Z)}[\mathcal{M}]_{(\rho,\xi)\to(\theta_1,\phi_1)}
\left[
\begin{array}{c}
\sin \theta_2 \cos\phi_2\\
\sin\theta_2 \sin\phi_2\\
\cos\theta_2
\end{array}
 \right]
\end{eqnarray}
where $[\mathcal{M}]_{(\rho,\xi)\to(\theta_1,\phi_1)}$ represents
the rotation matrix from the $(\rho,\xi)$ framework to the
$(\theta_1,\phi_1)$ framework and
$[\mathcal{M}]_{(\theta_1,\phi_1)\to(X,Y,Z)}$ is the rotation
matrix from the $(\theta_1,\phi_1)$ framework to the macroscopic
one. Expressions for these matrices are:
\begin{eqnarray}
[\mathcal{M}]_{(\rho,\xi)\to(\theta_1,\phi_1)}= \left[
\begin{array}{ccc}
\sin\rho\cos\xi&\cos\rho\cos\xi&-\sin\xi\\
\sin\rho\sin\xi&\cos\rho\sin\xi&\cos\xi\\
\cos\rho&-\sin\rho&0\\
\end{array}
 \right]
\end{eqnarray}
and:
\begin{eqnarray}
[\mathcal{M}]_{(\theta_1,\phi_1)\to(X,Y,Z)}= \left[
\begin{array}{ccc}
\sin\theta_1\cos\phi_1&\cos\theta_1\cos\phi_1&-\sin\phi_1\\
\sin\theta_1\sin\phi_1&\cos\theta_1\sin\phi_1&\cos\phi_1\\
\cos\theta_1&-\sin\theta_1&0\\
\end{array}
 \right]
\end{eqnarray}
The  fluorescence intensity emitted by the acceptor can then be
expressed~as:
\begin{eqnarray}
J_X(\Omega_2,\Omega_1,\rho,\xi)&=&K_1\,
\mu^2_{2X}(\theta_2,\phi_2,\rho,\xi,\theta_1,\phi_1)+K_2\,
\mu^2_{2Y}(\theta_2,\phi_2,\rho,\xi,\theta_1,\phi_1)\nonumber\\
&+&K_3\, \mu^2_{2Z}(\theta_2,\phi_2,\rho,\xi,\theta_1,\phi_1)\nonumber\\
J_Y(\Omega_2,\Omega_1,\rho,\xi)&=&K_2\,
\mu^2_{2X}(\theta_2,\phi_2,\rho,\xi,\theta_1,\phi_1)+K_1\,
\mu^2_{2Y}(\theta_2,\phi_2,\rho,\xi,\theta_1,\phi_1)\nonumber\\
&+&K_3\, \mu^2_{2Z}(\theta_2,\phi_2,\rho,\xi,\theta_1,\phi_1)
\end{eqnarray}
where the parameters $K_1$, $K_2$ and $K_3$ are defined in
Appendix A.

\subsection*{Appendix C: Polarization responses of SHG in a high NA microscope setup}
In the case of second harmonic generation, the expression of the
radiated field is similar to the fluorescence emission, requiring
mere replacement of the dipole moment $\vec{\mu}$ by the induced
second order nonlinear polarization~$\vec{P}^{(2)}$:
\begin{equation}
\vec{\mathcal{E}}^{\textrm{ SHG (radiated)}}(u,v)\propto\vec{k}
\wedge (\vec{P}^{(2)} \wedge \vec{k})
\end{equation}
with:
\begin{equation}
\vec{P}^{(2)}=\left[ \begin{array}{c} \chi^{(2)}_{\scriptscriptstyle{XXX}}\\ \chi^{(2)}_{\scriptscriptstyle{YXX}}\\
\chi^{(2)}_{\scriptscriptstyle{ZXX}}
\end{array}\right]  E_X^2  + \left[
\begin{array}{c} \chi^{(2)}_{\scriptscriptstyle{XYY}}\\ \chi^{(2)}_{\scriptscriptstyle{YYY}}\\ \chi^{(2)}_{\scriptscriptstyle{ZYY}} \end{array}\right] E_Y^2  +
2 \left[ \begin{array}{c} \chi^{(2)}_{\scriptscriptstyle{XXY}}\\
\chi^{(2)}_{\scriptscriptstyle{YXY}}\\
\chi^{(2)}_{\scriptscriptstyle{ZXY}}
 \end{array}\right]  E_X E_Y
\end{equation}
where the $\chi^{(2)}_{\scriptscriptstyle{IJK}}$ coefficients are
the tensor coefficients of the macroscopic nonlinear
susceptibility $\chi^{(2)}$. The radiated field
$\vec{\mathcal{E}}^{\textrm{ SHG(radiated)}}$ is then:
\begin{equation}
\vec{\mathcal{E}}^{\textrm{ SHG(radiated)}}(u,v)=\left[ \begin{array}{ccc} A_{XXX}&A_{XYY}&A_{XXY}\\
A_{YXX}&A_{YYY}&A_{YXY}\\
A_{ZXX}&A_{ZYY}&A_{ZXY}
 \end{array}\right]\left[
\begin{array}{c}   E_X^2  \\  E_Y^2
\\  E_X E_Y
\end{array}\right]
\end{equation}
where the $A_{IJK}$  coefficients  depend only on parameters
$(u,v)$. Taking into account the effect of the objective (see Eq.
(\ref{objective})), we can write:
\begin{equation}
\vec{\mathcal{E}}^{\textrm{ SHG}}(u,v)= [\mathcal{R}]\,\vec{\mathcal{E}}^{\textrm{ SHG (radiated)}}= \left[ \begin{array}{ccc} B_{XXX}&B_{XYY}&B_{XXY}\\
B_{YXX}&B_{YYY}&B_{YXY}\\
B_{ZXX}&B_{ZYY}&B_{ZXY}
 \end{array}\right]\left[
\begin{array}{c}   E_X^2 \\  E_Y^2
\\  E_X E_Y
\end{array}\right]
\end{equation}
where the $B_{IJK}$ coefficients depend only on the $(u,v)$
angles. As SHG signal is  emitted through a coherent process,
integration over all ($u,v$) angles in the objective aperture is
performed before taking the square of the emitted field. The total
collected field $\vec{\mathcal{E}}^{\textrm{ SHG}}_{\textrm{tot}}$
is then obtained by integration over the ($u,v$) angular
variables:
\begin{equation}
\vec{\mathcal{E}}^{\textrm{ SHG}}_{\textrm{tot}}=  \left[ \begin{array}{ccc} b_{XXX}&b_{XYY}&b_{XXY}\\
b_{YXX}&b_{YYY}&b_{YXY}\\
b_{ZXX}&b_{ZYY}&b_{ZXY}
 \end{array}\right]\left[
\begin{array}{c}   E_X^2  \\ E_Y^2
\\  E_X E_Y
\end{array}\right]
\end{equation}
where:
\begin{equation}
b_{IJK}=\int_0^{2\pi} \int_0^{\theta_{\textrm{obj}}} B_{IJK}(u,v)
\sin u\, \textrm{d}u\, \textrm{d}v
\end{equation}
Since the detected intensity is  proportional to
$\overline{|\vec{\mathcal{E}}^{\textrm{ SHG}}_{\textrm{tot}}|^2}$,
the expression of $I_X$ and $I_Y$ are therefore:
\begin{eqnarray}
I_{I=X,Y}^{\textrm{SHG}}&=&b_{IXX}^2\,   \overline{E_X^4}   +
b_{IYY}^2\, \overline{E_Y^4}
 + (2\,b_{IXX}\,b_{IYY} + b_{IXY}^2)\, \overline{E_X^2E_Y^2}\nonumber\\
 &+& 2\,b_{IXX}\,b_{IXY}\,  \overline{E_X^3E_Y } +
2\,b_{IYY}\,b_{IXY}\,  \overline{E_XE_Y^3}
\end{eqnarray}

\subsection*{Acknowledgments} The authors are grateful to Christelle
Anceau for helpful comments and discussions. They also thank
Fran\c{c}ois Treussart and Andr\'e Clouqueur for technical support
and Rolland Hierle for help with the photo-lithographic technique
and work facility in the clean-room.


\begin{thebibliography}{99}
\bibitem{sample} The microscope slides were carefully cleaned with acetone,
ethanol and ultra-pure water. We verified that undopped polymer
sample did not exhibit any measurable signal under two-photon
excitation.
\bibitem{frequence} The wavelength distribution in the
femtosecond pulses is neglegted in this model and only one
$\omega$ fundamental pulsation is considered.
\bibitem{gouy} The Gou\"y phase shift factor
contribution has been shown to be of crucial importance in the
case of SHG transmitted signals with distances along $Z$ longer
than the coherence length of the nonlinear process. In our
reflection configuration, the nonlinear interaction takes place in
the sample plane and the propagation length is lower than the
coherence length, therefore a possible $Z$ contribution factor
does not affect the propagation of the SHG signal. This factor
only brings a constant contribution to the SHG polarized signal
level, which has been quantified accounting for our 1.4 numerical
aperture, and seen to be negligible compared with the other
in-plane contributions.
\bibitem{angle} The angle between the two dipoles of absorption and emission for DCM
has been estimated separately using static fluorescence anisotropy
measurements in solution. The measured value is around
25$^{\circ}$, which does not significantly influence the results
in the model, compared with the other effects described in this
work.
\end{thebibliography}
\end{document}